\documentclass[runningheads]{llncs}
\usepackage[english]{babel}
\usepackage[utf8]{inputenc}

\makeatletter
\renewcommand*\l@section{\@dottedtocline{1}{1.5em}{2.3em}}
\makeatother

\usepackage[a4paper, left=25mm, right=25mm, top=30mm]{geometry}

\usepackage{fancyhdr}
\usepackage{multicol}
\usepackage{paracol}
\usepackage[labelfont=bf]{caption}
\usepackage{hyperref}
\usepackage{color}
\usepackage{changepage}
\usepackage{listings}
\usepackage{xspace}
\usepackage{textcomp}
\usepackage[dvipsnames]{xcolor}
\usepackage{wrapfig}

\usepackage{amsmath}
\usepackage{amssymb}

\usepackage{pgf}
\usepackage{pgfplots}
\usepackage{pgfplotstable}
\usepackage{tikz}
\pgfplotsset{compat=newest}
\usetikzlibrary{shapes,fit,decorations.pathmorphing}
\usepackage{siunitx}
\usepackage{makecell}
\usepackage{pifont}
\usepackage{bbding}
\usepackage{float}

\usepackage{todonotes}
\usepackage{soul}

\usepackage[linesnumbered]{algorithm2e}
\SetKwComment{Comment}{//}{}
\RestyleAlgo{ruled}

\usepackage{orcidlink}

\usepackage[T1]{fontenc}
\usepackage{graphicx}
\newcommand{\hlc}[2][yellow]{{%
    \colorlet{foo}{#1}%
    \sethlcolor{foo}\hl{#2}}%
}

\newcommand{\xmark}{\ding{53}}
\newcommand{\cmark}{\checkmark}

\usepackage{booktabs}
\usepackage[normalem]{ulem}

\begin{document}
\title{Noise Injection for Performance Bottleneck Analysis}
\titlerunning{Noise Injection for Performance Bottleneck Analysis}
%
\author{Aur\'{e}lien Delval\Envelope \inst{1,2} \orcidlink{0000-0002-2707-6940} \and
Pablo de Oliveira Castro \inst{2} \orcidlink{0000-0001-9007-6145} \and
William Jalby \inst{2} \orcidlink{0000-0002-4975-5469} \and
Etienne Renault\inst{1} \orcidlink{0000-0001-9013-4413} }
\authorrunning{A. Delval et al.}
%
\institute{SiPearl, France
\email{\{aurelien.delval, etienne.renault\}@sipearl.com} \and
Université Paris-Saclay, UVSQ, LI-PaRAD, France\\
\email{\{aurelien.delval, pablo.oliveira, william.jalby\}@uvsq.fr}}
\maketitle
\begin{abstract}
Bottleneck evaluation plays a crucial part in performance tuning of HPC applications, as it directly influences the search for optimizations and the selection of the best hardware for a given code. In this paper, we introduce a new model-agnostic, instruction-accurate framework for bottleneck analysis based on \textit{performance noise} injection. This method provides a precise analysis that complements existing techniques, particularly in quantifying unused resource slack. Specifically, we classify programs based on whether they are limited by computation, data access bandwidth, or latency by injecting additional noise instructions that target specific bottleneck sources. Our approach is built on the LLVM compiler toolchain, ensuring easy portability across different architectures and microarchitectures which constitutes an improvement over many state-of-the-art tools. We validate our framework on a range of hardware benchmarks and kernels, including a detailed study of a sparse-matrix–vector product (SPMXV) kernel, where we successfully detect distinct performance regimes. These insights further inform hardware selection, as demonstrated by our comparative evaluation between HBM and DDR memory systems. 
\end{abstract}

\setcounter{footnote}{0}


\section{Introduction} \label{section:introduction}

Modern High-Performance Computing (HPC) systems are increasingly adopting heterogeneous architectures to meet the growing demands for computational performance and energy efficiency. This heterogeneity occurs at several levels:  processing units level (CPUs, GPUs, FPGAs, and other accelerators), memory level (HBM, DDR, CXL, NVLink, and others),  microarchitectures (e.g. on AArch64: Cortex-M, Cortex-A,  Neoverse-N1,  Neoverse-V1,  Neoverse-V2) and SIMD extension level. All those combinations offer a wide range of optimization options for both the programmer and the compiler, the goal being to achieve full and balanced use of hardware resources (computing elements, memory bandwidth, and I/O). 

However, in practice, this balance is rarely achieved, as one resource often becomes a bottleneck, hence the need for tools to detect them. Once an imbalance is identified, various well-known optimizations can address the specific performance issues. For compute-bound codes, improvements can come from vectorization or replacing costly operations like divisions and square roots. For data-access-bound codes, optimizing data access patterns is a common approach. Similarly, for I/O-bound codes, performance gains can be achieved by optimizing access patterns or scheduling synchronization points more efficiently. While some general optimizations can provide quick performance boosts, the most effective strategies typically depend on the specific characteristics of the code and the underlying hardware. Therefore, detecting bottlenecks is essential for guiding performance optimization and selecting the most suitable hardware system for a given workload, including considerations such as microarchitecture and memory types.

To identify application bottlenecks, existing research primarily relies on abstractions and appro\-ximations. Static approaches like the roofline model and its derivatives~\cite{gavoille2024armprojection} analyze FLOPS and arithmetic intensity to estimate the potential for improving compute or bandwidth-bound codes, but often neglect cache or latency effects. Simulation or emulation-based methods~\cite{binkert2011gem5} are either complex to configure or oversimplify hardware details (NUMA effects, memory controller latency). These methods are useful for testing different hardware configurations and predicting how an application might behave with more cores or memory, or even on future hardware. However, they have two major drawbacks: (1) they require extensive computation time and dedicated hardware, and (2) abstracting certain hardware details can lead to misinterpretation of bottlenecks. To overcome these limitations, alternative approaches not relying on abstract performance models have been developed, such as performance-event sampling techniques, or decremental analysis that remove parts of the binary to identify bottlenecks~\cite{bendifallah2016decan}. Yet, to interpret measured data, these still rely on some level of abstractions or assumptions. Thus, they may fail to accurately capture performance bottlenecks in the same way as purely static methods.

To better identify imbalances in resource usage, this paper introduces a novel approach based on \textit{noise injection}. The key idea is to insert assembly instructions (or \textit{noise}) into performance-critical sections of the code to stress various bottlenecks. Thanks to the strong Out-of-Order (OoO) capabilities of modern hardware, injecting noise into a constrained resource will significantly impact performance, whereas doing so in an underutilized resource will have little effect. In section \ref{section:noise_methodology}, we define the \textit{absorption} metric which quantifies how much noise can be injected into a code without slowdown. This metric is used to quantify the usage of each bottleneck source. Unlike existing techniques, this method does not rely on heavy abstractions nor strong assumptions about hardware behavior. As a result, 1) we can precisely determine the root causes of bottlenecks, with an accuracy on the order of a single cycle; 2) the method does not require a detailed performance model and can be easily ported to new architectures. Our approach is designed to be both modular and flexible. First, different types of noise can be defined to stress various resources such as the Floating Point Unit (FPU) or the Load Store Unit (LSU), as detailed in Section~\ref{section:noise_methodology}. Additionally, our method is fully integrated into the LLVM compiler infrastructure. Section~\ref{section:expvalidation} demonstrates our method using hardware characterization benchmarks that target memory bandwidth, latency, and compute performance. These benchmarks also serve as a basis for comparing different systems in terms of microarchitectures and memory types by evaluating their resilience to noise and revealing unexpected architectural behaviors. Finally, Section~\ref{section:use-cases} presents a case study on a sparse matrix-vector product kernel, showing how our tool detects transitions between bottleneck phenomena that remain hidden with other benchmarking tools (reviewed in Section~\ref{section:state-of-the-art}).

\section{Description of the Noise Injection Methodology} \label{section:noise_methodology}

\subsection{Preliminary Definitions}

\begin{figure}[tbp]
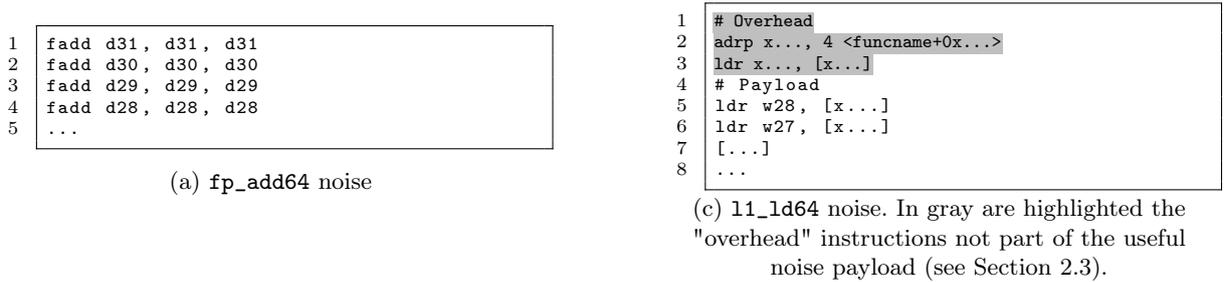

\centering
\begin{minipage}[t]{.45\textwidth}
    \vspace{0.3cm}
    \begin{lstlisting}[numbers=left,frame=lrtb, xleftmargin=18pt, basicstyle=\scriptsize\ttfamily]
fadd d31, d31, d31
fadd d30, d30, d30
fadd d29, d29, d29
fadd d28, d28, d28
...
    \end{lstlisting}
    \vspace{0.45cm}
    \vspace{-3ex}
    \centering(a) \lstinline{fp_add64} noise 
\end{minipage}
\hfill
\begin{minipage}[t]{.45\textwidth}
    \begin{lstlisting}[numbers=left,frame=lrtb, xleftmargin=18pt, basicstyle=\scriptsize\ttfamily]
<@\hlc[gray!50]{\# Overhead}@>
<@\hlc[gray!50]{adrp  x..., 4 <funcname+0x...>}@>
<@\hlc[gray!50]{ldr  x..., [x...]}@>
# Payload
ldr w28, [x...]
ldr w27, [x...]
[...]
...
    \end{lstlisting}
    \vspace{-1ex}
    \centering(c) \lstinline{l1_ld64} noise. In gray are highlighted the "overhead" instructions not part of the useful noise payload (see Section~\ref{subsection:payload}).
\end{minipage}
\captionof{figure}{Definition of basic AArch64 patterns, \lstinline{fp_add64} and \lstinline{l1_ld64}, which are useful to quantify compute and data-access bottlenecks.
\label{figure:noisetypes1}}
\vspace{-0.5cm}
\end{figure}

Let $\mathcal{I}$ be an Instruction Set Architecture (ISA). We use $\Sigma^*_{\mathcal{I}} $  to denote the set of finite sequences that we can build over $\mathcal{I}$, that is to say, the set of all possible assembly programs. A language $\mathcal{L}_{\mathcal{I}}$ is a subset of programs.
In this paper, we define the \textit{noise} language, $\mathcal{N}_{\mathcal{I}}$ which can be seen as a generator of assembly patterns. For the sake of simplicity, let us denote it by  $\mathcal{N}$. The language $\mathcal{N}$ is built as the union of multiple sub-languages called \textit{noise modes}. Several noise modes can be considered: \lstinline{fp_add64}, consisting in FP64 scalar add instructions; \lstinline{l1_ld64} consisting of scalar loads hitting the L1 cache, and \lstinline{memory_ld64} falling in memory; \lstinline{int64_add} consisting in integer scalar add instructions. Figure~\ref{figure:noisetypes1} depicts those noise modes in AArch64 ISA. The left-hand side (a) represents \lstinline{fp_add64} noise (i.e. \lstinline{fadd}s) while the right-hand side (b) represents \lstinline{memory_ld64} noise (i.e. \lstinline{ldr}s).

More complex noise modes are possible, with larger noise patterns or by combining multiple existing patterns. However, for all ISA and noise modes considered in this paper, the resulting languages are simple since they use alphabets of size one. For any such noise mode $\mathcal{N}_M \in \mathcal{N}$, let $\Sigma_{\mathcal{N}_M} = \{ n_{\mathcal{M}} \}$ be its alphabet, with $n_{\mathcal{M}}$ its corresponding \textit{noise pattern} (which we denote $n$ for simplicity). Thus the assembly snippets generated by $\mathcal{N}_M$ are $\mathcal{N}_M^* = \{ n^k, \forall \, k \in \mathbb{N} \}$, where $n^k$ is the word obtained after concatenating $k$ patterns $n$ (for $k$ = 0, we obtain $\epsilon$ the empty string). We call $k$ the \textit{noise quantity} of $n^k$.

\subsection{Resources and Saturation Phases}

The term \textit{hardware resources} in the context of CPUs refers to the fundamental components that enable processors, memory, and peripheral devices to interact effectively. In this paper, we mostly focus on the internal resources of the CPU (e.g. Arithmetic and Logic Unit, Vector Processing Units) and memory (e.g. Cache Hierarchy, Memory Management Unit). Depending on the microarchitecture, one can observe variations in terms of performance. Noise injection helps analyze performance bottlenecks by highlighting architectural variations. For example, on AArch64, the BF16 Dot Product (\lstinline{bfdot}) has a latency of 4 on Neoverse V1 and 5 on Neoverse V2, while the FP16 reduction (\lstinline{faddv}) latency decreases from 13 on Neoverse V1 to 8 on Neoverse V2.

By introducing increasing amounts of noise in these resources, both the original and noisy instructions begin to interact with each other. This interaction leads to performance degradation, eventually reaching a point where noise becomes dominant and fully saturates its target resource. To assess the availability of different CPU resources, we define \textit{absorption} as the number of additional instructions a code can handle before experiencing performance degradation. Naturally, this absorption capacity depends heavily on the target machine and the nature of the injected noise. In practice, up to three distinct phases may be observed, though the first two do not always appear.

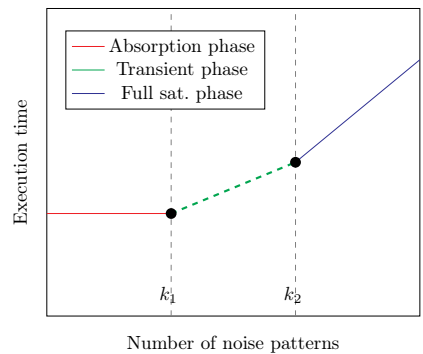
\begin{wrapfigure}[15]{r}{0.4\textwidth}
\vspace{-0.75cm}
\resizebox{0.35\textwidth}{!}
{
\begin{tikzpicture}[scale = 1]
    \begin{axis}[
    ylabel=Execution time,
    xlabel=Number of noise patterns,
    ylabel near ticks,
    xlabel near ticks,
    xmin=0, xmax=15,
    ymin=0, ymax=15,
    xticklabel=\empty,
    yticklabel=\empty,
    xtick style={draw=none},
    ytick style={draw=none},
    legend style={at={(0.05,0.8)},anchor=west}
    ]

    \draw[style=dashed, color=gray] (axis cs:5,\pgfkeysvalueof{/pgfplots/ymin}) -- (axis cs:5,\pgfkeysvalueof{/pgfplots/ymax});
    \draw[style=dashed, color=gray] (axis cs:10,\pgfkeysvalueof{/pgfplots/ymin}) -- (axis cs:10,\pgfkeysvalueof{/pgfplots/ymax});
    
    \draw[color=Red] (axis cs:0,5) -- (axis cs:5,5);
    \draw[color=Green, dashed,line width=0.40mm] (axis cs:5,5) -- (axis cs:10,7.5);
    \draw[color=Blue] (axis cs:10,7.5) -- (axis cs:15,12.5);

    \node[label={120:{}},circle,fill,inner sep=2pt] at (axis cs:5,5) {};
    \node[label={120:{}},circle,fill,inner sep=2pt] at (axis cs:10,7.5) {};

    \node[label={120:{$k_1$}}] at (axis cs:5.7,0) {};
    \node[label={120:{$k_2$}}] at (axis cs:10.7,0) {};

    \addlegendimage{no markers, color=Red}
    \addlegendimage{no markers, color=Green}
    \addlegendimage{no markers, color=Blue}
    \addlegendentry{Absorption phase }
    \addlegendentry{Transient phase}
    \addlegendentry{Full sat. phase}

    \end{axis}
\end{tikzpicture}
}\centering
\caption{Idealized model of codes' behavior when subject to increasing noise.}
\label{figure:idealmodel}
\end{wrapfigure}

First, an \textit{absorption phase}, during which performance is not affected at all by the noise. Injected instructions are filling cycles that were previously stalling, either because of bubbles in the pipeline or high latencies of some memory operations. Then, a \textit{transient phase}, where performance starts degrading. In the general case, the behavior of performance can become unpredictable and unstable, as it will be very dependent on the OoO implementation of the tested machine \cite{tomasulo1967ooo}. Finally, a \textit{saturation phase}, where injected noise becomes completely dominating compared to the original case. The system has reached its asymptotic behavior as run time starts increasing linearly, or at least in a consistent way. Figure~\ref{figure:idealmodel} shows an idealized representation of how a code responds to noise, highlighting the three phases just described. During the absorption phase, performance remains unchanged, appearing as a flat section in the plot. After $k_1$ patterns, performance starts suffering from injected noise. After $k_2$ patterns, the code is fully saturated and the system reaches asymptotic behavior (more details in Sec.~\ref{section:expvalidation}). To characterize the usage of a given resource, the \textit{absorption} metric is used. This corresponds to the value of $k_1$ in our model\footnote{When measuring experimental data, it is possible to automatize the computation of absorption by fitting the obtained series to this model.}. The main idea between this metric is that a \textit{lower value} indicates a \textit{saturated resource}, whereas a \textit{higher value} indicates an \textit{unsaturated resource}.

\subsection{Payload and Overhead} \label{subsection:payload}

One aspect requiring special attention is the side effects introduced by our noise, as it is important to ensure that it does not affect the semantics of the original code. Given $\mathcal{L}_c$ the language of the snippet of code and $\mathcal{N}_M$ the chosen noise mode. Without loss of generality, we consider here $\mathcal{L}_c$ to be a finite language. When injecting any quantity $k$ of noise into $\mathcal{L}_c$, we obtain $\mathcal{L}_R = \mathcal{L}_c \cup \mathcal{N}_M$ s.t any $s_r \in \mathcal{L}_R = s_1 . n^k . s_2$ with "$.$" the sequence  concatenation  symbol and $(s_1 . \epsilon . s_2) = (s_1  . s_2) \in \mathcal{L}_c$. Trivially for $k=0$,  the property holds since $n^k = \epsilon$ and $(s_1 . s_2) \in \mathcal{L}_c$. 
For arithmetic noise, one can consider a machine with an infinite number of registers (without loss of generality since the compilers can perform \textit{spilling} to conform to the limited number of registers of actual machines). Therefore the set of registers used per $n^k$, denoted $\mathcal{R}_n$ can differ from the set of registers used by both $s_1$ and $s_2$ (denoted by $\mathcal{R}_{s}$).
As a consequence, the evaluation of $s_r$ restricted to registers $\mathcal{R}_{s}$ will result in $s_1 . s_2$. Since $\mathcal{R}_n$ cannot impact other registers, the original semantics are not affected. In such a situation, similarly to what is done with partial order reduction in model-checking, one can build one representative execution, ignoring some others, with the same semantics. Notice that this scheme of proof also holds for other noise modes: one just has to ensure that there are no conflicts between the noise semantics and the code semantics.

In practice, there may be cases where register spilling can not be avoided. However, this will only happen when introducing a read-after-write (RAW) dependency, as opposed to a write-after-write (WAW) dependency. The phenomenon can also be mitigated by limiting the number of registers cycled by the noise pattern, as long as it is sufficient to not induce stalls between the noise instructions themselves. It is also possible to detect when spilling does happen by statically analyzing the generated assembly code. All in all, while register spilling is a constraint to be aware of, it can be limited and detected by the techniques mentioned above, and its likeliness depends on a wide range of criteria, such as the register pressure of the original code (already a potential performance issue by itself), the target microarchitecture, or the noise mode. Some noise modes might also have an inherent \textit{overhead} to set up the relevant noise instructions. This observation can be generalized by splitting injected instructions into two sets: \textit{payload} instructions -corresponding to the useful noise instructions to inject- and \textit{overhead} (whether they are spill instructions or inherent to the noise mode). The size of both sets of instructions can be computed by statically analyzing the code produced by the compiler. This permits evaluation of the \textit{quality} of the injection, ensuring that noise did not produce unexpected and significant side effects that may bias analysis. Note that when possible (e.g. when using previously untouched callee saved registers), the compiler can hoist overhead instructions out of the target loop, making its impact as small as possible.

\subsection{Renormalization of Absorption by Code Size}

Another important factor to consider is the impact of the noise quantity relative to the number of instructions in the original loop. Larger code sizes may indeed provide more opportunities for the OoO execution engine to utilize a noise instruction to fill a stall. This is particularly important when comparing different versions of the same code. The most obvious example of this is loop unrolling. This brings us to the definition of two methods for measuring absorption. $s_r = s_1.n^k.s_2$ is the assembly program obtained after injection of $k$ patterns of any noise mode $\mathcal{N}_M$. When only considering loop nests, $l_r$ can be defined as the substring of $s_r$ corresponding to the target loop body. Because $n^k$ is itself a substring of $l_r$, it is possible to write $l_r=l_1.n^k.l_2$. The \textit{relative payload} size is defined as 

\begin{equation}
    \widehat{P}(k) = \frac{k}{|l_1.l_2|}
\end{equation}

where $|l_1.l_2|$ denotes the size of $l_1.l_2$. Now, let $Abs_\mathcal{N}^{raw}(l_r)$ the absorption of loop $l_r$ for noise mode $\mathcal{N}_M$, expressed in number of raw noise patterns (that is the \textit{absolute} or \textit{raw} absorption). The \textit{relative absorption} is defined as 

\begin{equation}
    Abs_\mathcal{N}^{rel}(l_r) = \widehat{P}(Abs_{\mathcal{N}}^{raw}(l_r))
\end{equation}

Intuitively, this corresponds to the number of patterns that can be injected per instruction of the original loop body before reaching saturation.
Depending on the context, it may be more suitable to use either the absolute or relative definition of the absorption metric, the relative definition being particularly adapted to the study of loop mutations that greatly affect code size (such as unrolling or interleaving).


\section{Implementation of a Compiler-Level Injection Framework} \label{section:implementation}

Different approaches can be used to implement a framework for code modifications — specifically, for instruction injection. The first approach, binary patching, modifies compiler-generated binaries directly. Although successfully employed by tools such as MAQAO and MADRAS~\cite{valensi2014generic}, binary patching is challenging to port to new architectures; for example, MADRAS does not yet support AArch64. The second approach, dynamic code manipulation, is implemented by tools like Intel PIN~\cite{luk2005pin}, DynamoRIO~\cite{bruening2004dynamorio}, and Valgrind~\cite{nethercote2007valgrind}. While these tools are mature, they incur a run time overhead that is undesirable when precise performance measurements are required. 

\input{llvm_pipeline}

\subsection{Overview of the Tool} \label{subsection:tooloverview}

As a consequence, a compiler-level injection approach was chosen, building on the LLVM~\cite{lattner2004llvm} infrastructure. Although this confines the tool to a single toolchain, LLVM’s growing adoption in the HPC landscape and its support for distributing plugins as standalone middle-end passes make this limitation acceptable. Figure~\ref{fig:pipelinenoise} illustrates our LLVM-based injection workflow, which primarily consists of a plugin pass extending the compiler middle end. Our approach requires that the injected noise be both fine-grained and semantics-preserving, while also being robust against compiler optimizations. To meet these requirements, noise is injected via inline assembly at the LLVM Intermediate Representation (IR) level.  This allows writing the noise directly in assembly and specifying used registers as clobbered, ensuring that live and callee-saved registers are properly saved and restored. Using the \lstinline{volatile} qualifier guarantees that the noise is not optimized away by the compiler. Additionally, interferences with other optimization passes are avoided by performing the injection at the last optimization pass in the middle end. Since LLVM IR is target agnostic, the method is easily portable to other architectures, with the main effort being the development of noise pattern generators for each target ISA.

In addition, and to simplify the process for end-users, a high-level \textit{noise controller} tool that automates the noise injection pass on target applications was developed. This tool relieves users from the tedious task of manually invoking the noise injection pass and manages experiments by automatically varying noise quantities and modes. Users can specify target loop nests either by using a custom loop pragma —integrated into the LLVM frontends to easily test different noise modes— or through a configuration file, which allows the use of the noise injection plugin without modifying the LLVM frontend. Although noise is typically injected into the innermost loop for maximum sensitivity, our mechanism also permits selecting any loop level for both injected noise and timing probes.

To reduce experimentation time, the tool employs several strategies. First, to be able to noise and monitor multiple hot regions simultaneously, it is important to be able to time them individually. This is done by placing timing probes (calling a custom runtime library) around each loop or loop nest individually. Additionally, a clustering algorithm groups executions into performance classes, assuming similar run times indicate shared characteristics~\cite{deoliveira2018subsetting}; each class is then analyzed independently. Also, an online saturation detection method monitors run times and deviations, halting injection when noise effects become significant. To minimize rebuild times, the tool selectively updates source files containing target regions, avoiding full rebuilds and saving time in large applications.

Supporting OpenMP and MPI parallelism requires special care. Our runtime library and \lstinline{memory_ld64} noise mode require careful handling: the former uses a hashmap for timing samples, which could be problematic with concurrent threads, while the latter loads from a dedicated buffer in a chaotic pattern to minimize cache hits and prefetching. Both issues are resolved using Thread Local Storage (TLS), assigning threads/processes their own hashmaps or load buffers. When possible, timing probes are placed around OpenMP parallel regions to ensure the main thread submits hashmap entries. Future work includes selectively noising threads or processes to induce desynchronization~\cite{fiedor2012multithreadnoise}.

\subsection{Methodology for Characterizing an Application} \label{subsection:methodology}

Studying an application through the noise injection method first requires knowing which hot loops are interesting to target. This information can be obtained using a profiler. With no prior information about those loops' bottlenecks, running the application with one or a few different noise quantities is usually a time saver. This will give an idea of the sensitivity of the code to the noise. Codes that are very sensitive to noise and can only absorb a few instructions -if any- are usually those that exhibit bottlenecks at the CPU core level. This is typically the case with compute-bound codes. On the other side of the spectrum, codes that can absorb very large quantities of noise (such as a few dozen instructions) are those that have some data access bottlenecks at the cache or memory levels, whether this is due to bandwidth or latency. 
Our experiments, on both mini-applications and real ones -and on various architectures-, show that values around 20 or 30 FP or L1 instructions are a good starting point, as this typically roughly corresponds to the tipping point between the two categories. 
Once the general sensitivities of the noise regions have been established, one can start running the code through increasing noise quantities following the process described in the previous section. 
For loops that appear robust to noise, it is usually preferable to use a step of 5 or even 10 instructions between successive noise payloads. The obtained absorption value should give indications about the general behavior of each target loop for further studies.
\section{Validation on Reference benchmarks }
\label{section:expvalidation}
\pgfplotscreateplotcyclelist{abscolors}{
Red,
Green,
Blue
}

\begin{figure}[tbp]
\begin{minipage}[t]{0.45\textwidth}
\begin{center}
\begin{tikzpicture}
    \begin{axis}[
    ylabel=Avg. CPU cycles,
    xlabel=Num. of instructions,
    scaled ticks=false,
    yticklabel={
              \num[
                round-mode=figures,
                scientific-notation=fixed,
                fixed-exponent=5
              ]{\tick}
    },
    yticklabel style = {
    font=\small
    },
    ylabel near ticks,
    xlabel near ticks,
    xmin=0, xmax=50,
    scale=0.75,
    cycle list name=abscolors
    ]
    \addplot table [col sep=comma, x=n, y=mean, mark=none]{matprod_g3_O0_arithmetic_fp64.csv};
    \addplot table [col sep=comma, x=n, y=mean, mark=none]{matprod_g3_O0_data_l1.csv};

    \draw[style=dashed, color=Red] (axis cs:11,\pgfkeysvalueof{/pgfplots/ymin}) -- (axis cs:11,\pgfkeysvalueof{/pgfplots/ymax});
    \draw[style=dashed, color=Green] (axis cs:1,\pgfkeysvalueof{/pgfplots/ymin}) -- (axis cs:1,\pgfkeysvalueof{/pgfplots/ymax});
    
    \addlegendimage{no markers, color=Red}
    \addlegendimage{no markers, color=Green}
    \addlegendentry{\lstinline{fp_add64}}
    \addlegendentry{\lstinline{l1_ld64}}
    \end{axis}
\end{tikzpicture}
(a) Absorption when compiling with \lstinline{-O0}
\end{center}
\end{minipage}
\hfill
\begin{minipage}[t]{0.45\textwidth}
\begin{center}
\begin{tikzpicture}
    \begin{axis}[
    ylabel=Avg. CPU cycles,
    xlabel=Num. of instructions,
    scaled ticks=false,
    yticklabel={
              \num[
                round-mode=figures,
                scientific-notation=fixed,
                fixed-exponent=4
              ]{\tick}
    },
    ylabel near ticks,
    xlabel near ticks,
    xmin=0, xmax=18,
    scale=0.75,
    cycle list name=abscolors
    ]
    \addplot table [col sep=comma, x=n, y=mean, mark=none]{matprod_g3_O3native_arithmetic_fp64.csv};
    \addplot table [col sep=comma, x=n, y=mean, mark=none]{matprod_g3_O3native_data_l1.csv};
    
    \draw[style=dashed, color=Red] (axis cs:0.35,\pgfkeysvalueof{/pgfplots/ymin}) -- (axis cs:0.35,\pgfkeysvalueof{/pgfplots/ymax});
    \draw[style=dashed, color=Green] (axis cs:0.2,\pgfkeysvalueof{/pgfplots/ymin}) -- (axis cs:0.2,\pgfkeysvalueof{/pgfplots/ymax});
    
    \addlegendimage{no markers, color=Red}
    \addlegendimage{no markers, color=Green}
    \addlegendentry{\lstinline{fp_add64}}
    \addlegendentry{\lstinline{l1_ld64}}
    \end{axis}
\end{tikzpicture}
(b) Absorption when compiling with \lstinline{-O3 -mcpu=native}
\end{center}
\end{minipage}

\caption{Absorption of a matrix product example with different compiler flags. Figure (a) absorbs 11 instructions, whereas in Figure (b), a single noise instruction increases running time demonstrating that the compiler optimizations effectively eliminate the performance bottleneck.
\label{figure:matprod}}
\vspace{-0.5cm}
\end{figure}
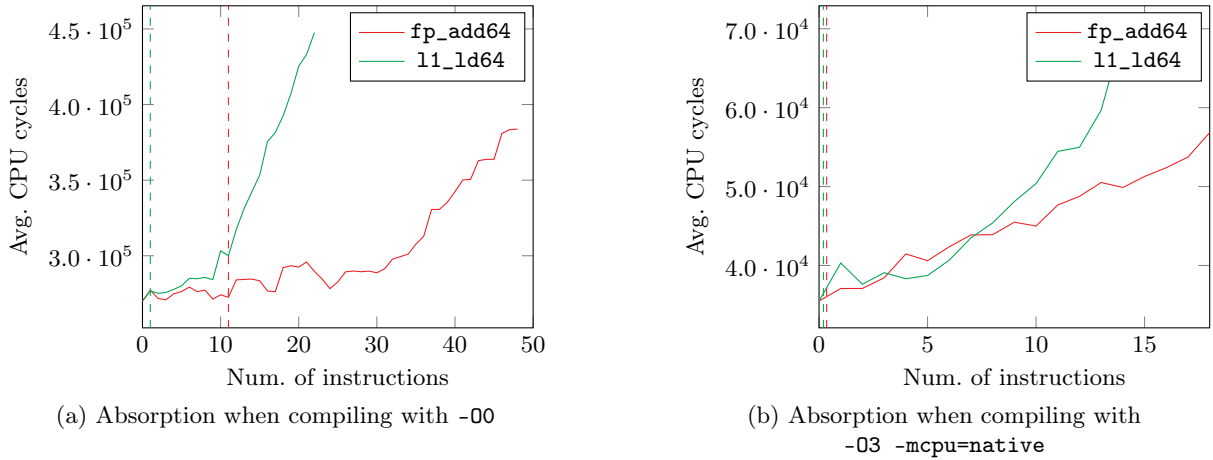

\subsection{Introductory Example: Dense Matrix Product}

Figure~\ref{figure:matprod}(a) demonstrates the noise injection pass on a simple matrix product example, running on an Amazon Graviton 3, and compiled with \lstinline{clang -O0}. The amount of noise is increased progressively to observe its impact. Under \lstinline{fp_add64} noise, the code appears quite robust, being completely unaffected by up to 11 instructions. Under \lstinline{l1_ld64} noise, however, performance degrades instantly. This clearly indicates that the code is data-access-bound. This is because the code was purposefully built in \lstinline{-O0}: in LLVM, this does not include the \lstinline{mem2reg} optimization pass, which promotes memory accesses to register accesses. As a consequence, the resulting assembly code is clogged with unnecessary load and store instructions that saturate the LSU while leaving the FPU mostly unused. Figure~\ref{figure:matprod}(b) presents the same experiment with \lstinline{-O3 -mcpu=native}, exploiting hardware resources much more efficiently.

\subsection{Validation Using Hardware Characterization Benchmarks} \label{subsecction:hardwarevalid}

We validate our noise injection methodology using established hardware characterization benchmarks: STREAM\footnote{\url{https://www.cs.virginia.edu/stream/ref.html}} for memory bandwidth, lat mem rd from LMBench\footnote{\url{https://lmbench.sourceforge.net/}} for memory latency, and Coral HACCmk\footnote{\url{https://asc.llnl.gov/coral-benchmarks\#haccmk}} for compute-bound operations. This approach serves two purposes: it confirms our tool's ability to detect and quantify specific hardware bottlenecks, and it provides practical insights for application developers through cross-architecture comparisons using the absorption metric.


\begin{figure}[tbp]
\centering
\begin{tikzpicture}[scale = 0.8]
    \begin{axis}[
    width=16cm,
    height=5cm,
    x tick style = transparent,
    restrict y to domain*=0:35,
    ymax=30,
    every axis plot post/.style={/pgf/number format/fixed},
    ybar=3pt,
    bar width=12pt,
    ymajorgrids = true,
    ylabel = Raw absorption,
    x tick label style  = {text width=2cm,align=center},
    symbolic x coords={a. STREAM (one core), b. STREAM (all cores),  c. lat mem rd (one core), d. HACCmk (one core)},
    xtick distance=1,
    xtickmin={a. STREAM (one core)},
    xtickmax={d. HACCmk (one core)},
    scaled y ticks = false,
    ymin=0,
    visualization depends on=rawy\as\rawy, 
    after end axis/.code={ 
            \draw [ultra thick, white, decoration={snake, amplitude=1pt}, decorate] (rel axis cs:0,1.05) -- (rel axis cs:1,1.05);
        },
    nodes near coords={%
            \pgfmathprintnumber{\rawy}
        },
    axis lines*=left,
    legend cell align=left,
    legend style={
        at={(1.3,0)},
        anchor=south east,
        column sep=1ex
    },
    clip=false
    ]

    \addplot[style={Red,fill=Red,mark=none}]
    coordinates {
        (a. STREAM (one core), 0.0)
        (b. STREAM (all cores), 65.0)
        (c. lat mem rd (one core), 250.0)
        (d. HACCmk (one core), 0.0)
    };

    \addplot[style={Green,fill=Green,mark=none}]
    coordinates {
        (a. STREAM (one core), 1.0)
        (b. STREAM (all cores), 26.0)
        (c. lat mem rd (one core), 240.0)
        (d. HACCmk (one core), 13.0)
    };
    
    \addplot[style={Blue,fill=Blue,mark=none}]
    coordinates {
        (a. STREAM (one core), 0.0)
        (b. STREAM (all cores), 0.0)
        (c. lat mem rd (one core), 15.0)
        (d. HACCmk (one core), 0.0)
    };
    
    \legend{\lstinline{fp_add64}, \lstinline{l1_ld64}, \lstinline{memory_ld64}}
    
    \end{axis}
\end{tikzpicture}
\caption{Raw absorptions for the three hardware characterization benchmarks on Amazon Graviton 3.}
\label{figure:validationbenchmarks_g3}
\vspace{-0.5cm}
\end{figure}
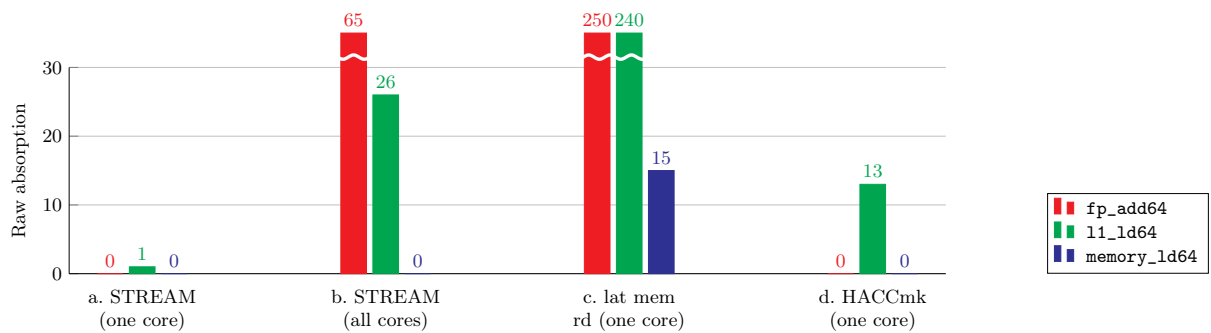

\noindent\textbf{Validation on Graviton 3.}
Figure~\ref{figure:validationbenchmarks_g3} shows results from all three benchmarks executed on an Amazon Graviton 3 system, with one scalar element loaded per iteration in both STREAM and lat mem rd.
\begin{itemize}
    \item For the \textbf{STREAM} benchmark, the sequential run (a.) results in low absorption values that reflect core-level limitations. In contrast, the parallel execution (b.) stresses the full memory bandwidth of the socket, causing loads to stall longer and enabling the absorption of many more noise instructions. Notably, while large quantities of \lstinline{fp_add64} and \lstinline{l1_ld64} noise are absorbed, \lstinline{memory_ld64} noise is not. To ensure that the absence of absorption in \lstinline{memory_ld64} noise is due to bandwidth saturation rather than an imbalance between injected noise and loop body size, we repeated the experiment after unrolling the code (see Table~\ref{table:aarch64comp}). These stable results confirm that the lack of absorption in \lstinline{memory_ld64} noise is a consequence of bandwidth saturation. 
    \item The \textbf{lat mem rd} benchmark reinforces these findings while highlighting an important distinction: unlike STREAM, it absorbs substantial \lstinline{memory_ld64} noise. This difference occurs because lat mem rd stalls first from dependencies between consecutive loads rather than bandwidth constraints. The ability to absorb up to 15 \lstinline{memory_ld64} instructions differentiates latency-bound from bandwidth-bound operations.
    \item The compute-bound \textbf{HACCmk} benchmark exhibits absorption only in \lstinline{l1_ld64}, with no measurable absorption in \lstinline{fp_add64}. This aligns with expectations that it saturates compute resources.
\end{itemize}

In summary, these results demonstrate that the noise injection methodology effectively captures all three types of bottlenecks — memory bandwidth, memory latency, and compute. By providing a nuanced, instruction-level absorption metric, this approach offers insights that go far beyond direct performance measures alone, thereby enabling more precise and targeted performance optimizations.

\noindent\textbf{Comparison of AArch64 and x86 Systems.} While Figure~\ref{figure:validationbenchmarks_g3} shows noise absorption on a specific (Graviton3) system,  Table~\ref{table:aarch64comp} provides a wider analysis by comparing performance and absorption metrics across three AArch64 systems - Ampere Altra (Neoverse N1), Amazon Graviton 3 (Neoverse V1), Nvidia Grace (Neoverse V2) - and two x86 systems - Sapphire Rapids with and without HBM.

\begin{itemize}
    \item For \textbf{STREAM} (bandwidth-bound benchmark), absorption inversely correlates with performance. Systems with more load stalls and lower performance exhibit higher absorption, confirming that stalled cycles have potential for being exploited. This is especially true for the x86 Sapphire Rapids machine where the only difference is the memory type. For this machine, we observe slightly more FP absorption with DDR, consistent with its lower performance. Variations can also be observed across AArch64 microarchitecture generations. The transition from N1 to V1 introduces a more advanced out-of-order execution engine (the pipeline core size increasing from 8 to 15), leading to better performance and increased absorption. A further evolution from V1 to V2 brings a slight performance improvement, which in turn reduces the amount of possible absorption. From a perspective of fine analysis of microarchitectural changes, our approach therefore exhibits interesting metrics.
    \item For \textbf{lat\_mem\_rd} (latency-bound benchmark), similar trends can be observed: absorption inversely correlates with performance. As we move across AArch64 microarchitecture generations, the amount of noise that can be injected increases. This reflects the growing complexity of newer chips, featuring larger network-on-chip (NoC) designs and more advanced memory systems (DDR4 vs. DDR5) with higher latency. On Sapphire Rapids, absorption capacity plateaus, highlighting the well-known NoC saturation~\cite{mccalpin2023noc} issue in this architecture. While this may limit its usefulness for characterization, exploring access patterns could provide insights into a broader range of real-world applications.
    \item In the compute-bound \textbf{HACCmk} benchmark, we observe microarchitectural differences. Graviton 3 (V1) and Nvidia Grace (V2) behave similarly, absorbing significant \lstinline{fp_add64} noise with minimal \lstinline{l1_ld64} absorption. In contrast, Ampere Altra (N1) shows no measurable absorption, suggesting either more balanced resource usage or a frontend/dispatch bottleneck. The same goes for Sapphire Rapids, where the compiler was able to vectorize the code, thus enabling further optimizations. This highlights how microarchitectural implementations can produce markedly different behaviors across or even within the same instruction set architectures.
    
\end{itemize}

Our absorption metric thus provides fine-grained insights into hardware resource usage beyond what traditional performance metrics reveal, offering a valuable tool for system selection and optimization. In particular, unicore measurements can be extremely misleading in driving optimization search: our experiments show that small code changes will often have an impact on unicore performance (as demonstrated by the low absorption values), but will often be overshadowed by much stronger data-access effects when scaling to multicore. This will be shown again in Figure~\ref{figure:spmxvheatmap} on the SPMXV kernel example. 

\begin{table}[t]
\begin{center}

\captionof{table}{Raw absorptions on different systems. All benchmarks were compiled with LLVM 18.1.8. \textit{($*$)~Code for this benchmark/noise combination was unrolled to better highlight resources saturation.}}

{\setlength{\tabcolsep}{5pt} 
\begin{tabular}{@{}lccccc@{}}
\toprule

\textbf{Machine:} & Ampere Altra & Graviton 3 & Grace & \multicolumn{2}{c}{Sapphire Rapids}\\ 
Microarchitecture: & \textbf{Neoverse} & \textbf{Neoverse} & \textbf{Neoverse} & \multicolumn{2}{c}{\textbf{Golden Cove}}\\
 & \textbf{N1} & \textbf{V1} & \textbf{V2} \\
Cores count: & 80 & 64 & 72 & \multicolumn{2}{c}{40}\\ 
Base Core frequency (GHz): & 3.0 & 2.6 & 3.2 & \multicolumn{2}{c}{2.2}  \\
Socket count:& 2 & 1 & 2 & \multicolumn{2}{c}{2} \\ 
Memory Type: & DDR & DDR & DDR & DDR & HBM\\
\midrule
\midrule
\textbf{STREAM (max core count)}\\
Performance (GB/s): & 168 GB/s & 262 GB/s & 381 GB/s & 211 GB/s & 541 GB/s \\
FP/L1/mem$*$ abs. & 47/27/0 & 65/26/0 & 21/16/0 & 80/80/0 & 24/21/0 \\

\midrule

\textbf{lat\_mem\_rd (1 core)}\\
Performance (ns): & 87.7 ns & 118 ns & 153 ns & 92 ns & 122 ns\\
FP/L1/mem abs. $\approx$ & 90/20/2 & 250/240/15 & 300/300/16 & 270/180/18 & 270/180/18\\

\midrule

\textbf{HACCmk (1 core)}\\
Performance (s): & 6.02 s & 9.85 s & 3.65 s & \multicolumn{2}{c}{4.25 s} \\
FP/L1/mem abs. & 0/0/0 & 0/13/0 & 0/9/0 & \multicolumn{2}{c}{0/0/0} \\

\bottomrule 

\end{tabular}
\medbreak
\label{table:aarch64comp}
}
\end{center}
\vspace{-1cm}
\end{table}

\section{Comparison with other State-of-the-Art Characterization Tools}

\subsection{Existing Methods for Detecting Performance Bottlenecks.}

\label{section:state-of-the-art}

The key idea developed in this paper is to propose a framework to insert assembly instructions into performance-critical sections. As a consequence, we can characterize, with a cycle level precision, how much hardware resources are saturated.
In the literature, other methods for performance and bottleneck analysis exist but have limitations. Table~\ref{table:toolscomparison} summarizes the strengths and weakness of these approaches which fall into three categories: dynamic, static, or modification. Each technique is evaluated w.r.t. portability (criteria 1), hardware specificity (criteria 2), robustness (criteria 3), low-side effects (criteria 4),  interpretability (criteria  5), and execution cost (criteria 6). A closer look shows that our approach meets most of the criteria compared to existing work. Indeed: the LLVM-based implementation ensures easy portability (1); the method is purely based on measuring execution time and does not make any abstraction through performance models (2); it applies to the vast majority of codes without specific restrictions, with support for OpenMP and MPI parallelism (3); overhead on top of noise payloads is reduced to the minimum and rarely necessary (4); the absorption metric is measured in instructions, which gives an intuitive evaluation of available resources (5). However, despite efforts to reduce the cost of this process, it does require re-building and re-running the target application multiple times, which can be a long process depending on the use case (6). The following subsections further detail other methods and tools and evaluate them with regard to the six criteria.

\noindent\textbf{Dynamic Analysis.}
A common approach to hardware performance monitoring is using standard tools like Linux's \lstinline{perf} to access Hardware Performance Monitoring Units (PMU). These counters provide insights for cache hits, cache misses, stalls, IPC, and more. 
However, these counters are often poorly documented (depending on the microarchitecture) and trigger security problems\footnote{It requires adjusting the \lstinline{perf_event_paranoid} kernel setting on Linux systems and can give insights about applications that run in parallel in dockerized/virtual machine contexts.}. Even if some effort has been made to create unified interfaces (such as PAPI~\cite{mucci1999papi}) or analysis methods (Intel Top-Down~\cite{ahmad2014topdown}), this issue persists. More fundamentally, interpreting hardware event data can be challenging~\cite{moseley2011hpm} even for advanced users, making it difficult to diagnose performance issues (criteria 5). Lastly, the roofline model and its derivatives measure a code's FLOPS, combined with an evaluation of arithmetic intensity. Comparing these to the peak performance of the machine provides a quick way to estimate if a code is bound by the memory bandwidth or by the peak compute performance of the machine. However, this family of models neglects architectural aspects that are also bottleneck sources, such as memory latency, cache hierarchies, different NUMA nodes, ...

\begin{figure}[tbp]
\begin{center}
\captionof{table}{Comparison of bottleneck analysis methods. "\cmark": the method meets the criteria; "\xmark": the method does not meet the criteria; "$\backsim$": the method meets the criteria to some extent. \textit{(*) Decremental analysis is evaluated on the basis of DECAN's implementation.}}
\begin{tabular}{ l||c||c c c c c c }
\hline 
\textbf{Method}  & Category 
& Portable & HW specific & Robust & Low side-effects & Interpretable & Fast \\ \hline
\textbf{PMU} & Dynamic & $\backsim$ & \cmark & \cmark & \cmark & \xmark & $\backsim$ \\
\textbf{Roofline model} & Dynamic & \cmark  & \xmark & \xmark & N/A & \xmark & $\backsim$ \\
\textbf{Code analyzers} & Static & \xmark & $\backsim$ & \xmark & N/A & \cmark & \cmark \\
\textbf{Decremental analysis\textit{(*)}} & Modification & \xmark & \cmark & $\backsim$ & \xmark & \cmark & $\backsim$ \\
\textbf{[this paper] Noise Injection } &Modification & \cmark & \cmark & \cmark & \cmark & \cmark & \xmark \\\hline
\end{tabular}
\label{table:toolscomparison}
\end{center}
\vspace{-1cm}
\end{figure}

\noindent\textbf{Static Analysis.}
An alternative approach is to simulate the performance of assembly code snippets. One such approach is the ECM model~\cite{hofmann2015ecm}, which provides performance estimation by combining both an in-core and a data transfer model. However, it does make some assumptions by neglecting cache and latency effects, making it similar to the roofline model. Other tools such as LLVM-MCA\footnote{\url{https://llvm.org/docs/CommandGuide/llvm-mca.html}}, 
uiCA~\cite{abel22uica} and MAQAO CQA~\cite{charif2014cqa} perform in-depth code analysis at the core level. CQA, in particular, is extended by UFS~\cite{palomares2016ufs}, which simulates execution on an idealized out-of-order (OoO) machine. Sniper generalizes this approach to multicore scenarios~\cite{carlson2011sniper} and Gus~\cite{pompougnac2024gus} allows the modification of microarchitectural constants to artificially tune available resources. Such static models offer the advantage of being fast and collecting critical metrics like port occupancy and stalls. However, they have several limitations. First, they make strong assumptions about memory accesses (limited by bandwidth or hitting L1 cache), which can lead to severely biased results when latency becomes the limiting factor or when the actual L1 hit rate is low (criteria 3). Second, porting these models to new architectures is challenging, as it requires fine-tuning with numerous microarchitecture-dependent constants, potentially taking weeks of expert work (criteria 1). In contrast, our approach only requires basic knowledge about the target instruction set to port the existing noise modes (although defining more advanced noise modes could require finer knowledge). Finally, static models struggle to accurately capture complex phenomena like instruction reordering and branch prediction, making them less reliable substitutes for real performance measurements (criteria 2). 

It is also worth mentioning that various machine learning techniques have been applied to the problem of performance prediction~\cite{wu2022ml}.

\noindent\textbf{Modification-Based Approaches.}
This category comprises methods that modify the target program to infer performance characteristics from the modified version, such as the noise injection method presented in this paper. 
In particular, it is quite similar to differential analysis, a method implemented by MAQAO's DECAN module~\cite{koliai2013decan}. DECAN generates different variants of an original binary, each corresponding to a potential bottleneck source for which instructions have been removed. By comparing the performances of the original and modified program versions, it is possible to infer which resource was the main bottleneck in the original code. Noise injection performs similar modifications to deduce performance bottlenecks. However, it adopts an \textit{incremental} approach, while DECAN is \textit{decremental}. While both can be complementary, removing instructions from the original code can have several harmful side effects (criteria 4). The most obvious one is that this completely breaks the code's semantics. DECAN solves this by keeping a copy of the original loop body and executing both versions at each iteration, so as to preserve the original control flow and overall behavior. More problematic, deleting instructions will have unpredictable side effects on the remaining ones. First, this may remove inter-instruction dependencies. Second, those deletions will free the tested bottleneck resource and all shared ones (reordering buffers, reservation stations), allowing the remaining instructions to "spread" on all available resources more easily. Finally, the implementation of differential analysis made by DECAN relies on MAQAO's MADRAS module for static binary patching~\cite{valensi2014generic}, making it hardly portable (criteria 1). The beginning of Section~\ref{section:implementation} goes into further detail on the topic of code modification. 

\noindent\textbf{Discussion.}
Some tools also use the idea of noise injection to evaluate programs' resilience, albeit in a different way and goal. The HPAS~\cite{ates2019hpas} and GREMLINS~\cite{maiterth2015gremlins} tools allow users to generate a variety of external system noises, whether that is by consuming CPU time, creating cache or memory contention. The goal is to evaluate programs' resilience to OS noise. Tools like REFINE~\cite{georgakoudis2017refine} use a similar approach to ours by injecting "internal" noise at the compiler level. However, they focus on evaluating resilience to soft errors, such as bitflips.

\subsection{Comparison with MAQAO DECAN}

\begin{figure}[tbp]
\begin{center}
    \captionof{table}{Comparison of DECAN and noise injection for FP/LS bottleneck analysis. In DECAN, a lower metric indicates a saturated resource, while in noise injection a lower absorption value signals a bottleneck.}
    \begin{tabular}{p{3cm} p{6cm} p{0.5cm} p{6cm}}
        \toprule
        \textbf{Scenario} & \textbf{DECAN (Decremental)} && \textbf{Noise Injection (Incremental)} \\
        \midrule
        1) Compute-bound   
        & LS variant runs significantly faster than FP variant (LS saturation is low, FP saturation is high) 
        && FP absorption is low, while LS absorption is high \\
        \midrule
        2) Data-bound      
        & FP variant runs significantly faster than LS variant (FP saturation is low, LS saturation is high) 
        && LS absorption is low, while FP absorption is high \\
        \midrule
        3) Full Overlap    
        & Both FP and LS variants run close to the original reference (both saturations are high) 
        && Both FP and LS absorptions are very low \\
        \midrule
        4) Limited Overlap 
        & Both FP and LS variants run significantly faster than the reference (both saturations are low) 
        && Ambiguous absorption levels (moderate) indicating strong interdependencies \\
        \bottomrule
    \end{tabular}
    \label{tab:decancomparison}
\end{center}
\vspace{-1cm}
\end{figure}

The previous section highlights some similitude with MAQAO DECAN~\cite{bendifallah2016decan,koliai2013decan}. On the one hand, DECAN uses a subtractive approach (it removes instructions). On the other, our approach is additive (it adds instructions). This section aims to clarify the main differences. First, it should be noted that both approaches have close strategies to stress hardware resources: floating-point arithmetic can be stressed by any \lstinline{fp_[...]} noise, while DECAN has a \texttt{FP} variant for that (keeping only FP instructions). Similarly, load/store is stressed through \lstinline{l1_ld64} or \lstinline{memory_ld64} noises, while DECAN has a \texttt{LS} variant. Despite these similarities, our approach is more fine-grained since one can inject any type and quantity of noise, whereas DECAN deletes all instructions of some given types. DECAN defines its \textit{saturation} metric as 

\begin{equation}
    Sat(\text{VAR}) = \frac{T(\text{VAR})}{T(\text{REF})}
\end{equation}

with $T(\text{VAR})$ and $T(\text{REF})$ designating the execution times of the modified and reference codes respectively. Therefore in DECAN, a lower metric (i.e., a variant running faster than the reference) signals that the removed resource was saturated, whereas in our method a lower absorption value indicates that only a few injected instructions are needed to degrade performance, thereby identifying a bottleneck.

\pgfplotscreateplotcyclelist{rescalecolors}{
Red,
Green,
Blue
}
\begin{wrapfigure}[19]{r}{0.4\textwidth}
\begin{tikzpicture}
    \begin{axis}[
    ylabel=Time {[}sec.{]},
    xlabel=Normalized noise quantity $\widehat{N}$,
    ylabel near ticks,
    xlabel near ticks,
    xmax=0.35,
    scale=0.7,
    legend style={at={(0.018,0.86)},anchor=west},
    no markers,
    cycle list name=rescalecolors
    ]
    \addplot table [col sep=comma, x=nRescaled, y=mean, mark=none]{livermore_fp.csv};
    \addlegendentry{\lstinline{fp_add64}}
    \addplot table [col sep=comma, x=nRescaled, y=mean, mark=none]{livermore_ls.csv};
    \addlegendentry{\lstinline{l1_ld64}}
    \end{axis}
\end{tikzpicture}
\caption{\lstinline{livermore_livermore:lloops.c_1351} kernel with FP arithmetic and L1 noise. It exhibits little to no absorption with similar behaviors between noise modes, hinting at a frontend bottleneck.}
\label{figure:livermore}
\end{wrapfigure}
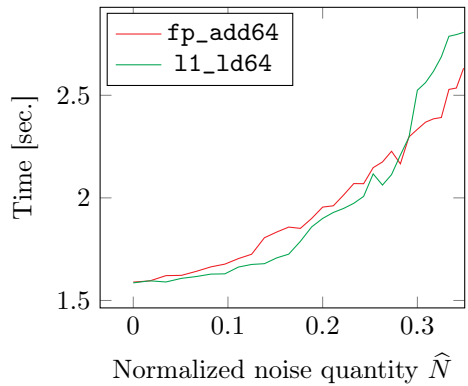

Table~\ref{tab:decancomparison} summarizes the expected behavior of DECAN versus our noise injection method with regard to their respective metrics. The limited overlap scenario of case 4) can not be fully understood by DECAN only, as it can arise as a consequence of two situations: either there are heavy dependencies between the FP and LS instructions flows, or the code has a shared microarchitectural bottleneck at the frontend level that disappears after removing instructions. Noise injection on the other hand is capable of distinguishing them. Because of our incremental approach, we do not alter the original pressure on dispatch ports, allowing us to distinguish both cases.

\noindent\textbf{Experimental Comparison.}
This section illustrates the above discussion on an example picked from the LORE loop repository~\cite{chen2017lore}. Because DECAN is available only on x86 (cf. the "Portability" criteria, mentioned in the previous section), this experiment was run on an Intel Xeon Gold 6254 CPU. 
Figure~\ref{figure:livermore} demonstrates how combining our noise injection with DECAN's differential analysis enables fine-grained performance investigations. The function \lstinline{livermore_livermore:lloops.c_1351} comprises two major dependency channels of FP computations using identical input values, resulting in a relatively high arithmetic intensity of 0.22 FP operations per loaded byte, as reported by CQA. At first glance, one might expect this code to be compute-bound. This expectation is supported by DECAN, which yields $Sat_{FP}=0.81$ and $Sat_{LS}=0.12$, suggesting an FP bottleneck. However, the absorption values tell a different story: as shown in the figure, both $Abs_{FP}^{rel}$ and $Abs_{LS}^{rel}$ approach zero with similar trends, a condition that might indicate full utilization and overlapping of the two resources (full saturation, case 3). DECAN, however, has already ruled out this full saturation scenario, implying instead a frontend bottleneck. Furthermore, MAQAO's CQA and UFS static code analyzers corroborate this interpretation by reporting the front end as a bottleneck —estimating an overhead of 0.75 cycles per iteration— although its subtractive approach caused DECAN to overlook it. Integrating differential analysis with noise injection uncovered a second critical bottleneck in the code. This case is particularly challenging because the overlapping bottleneck sources would have been missed by DECAN alone, potentially guiding programmers toward suboptimal optimizations.
\section{Use Case: Analysis of the SPMXV Kernel}
\label{section:use-cases}

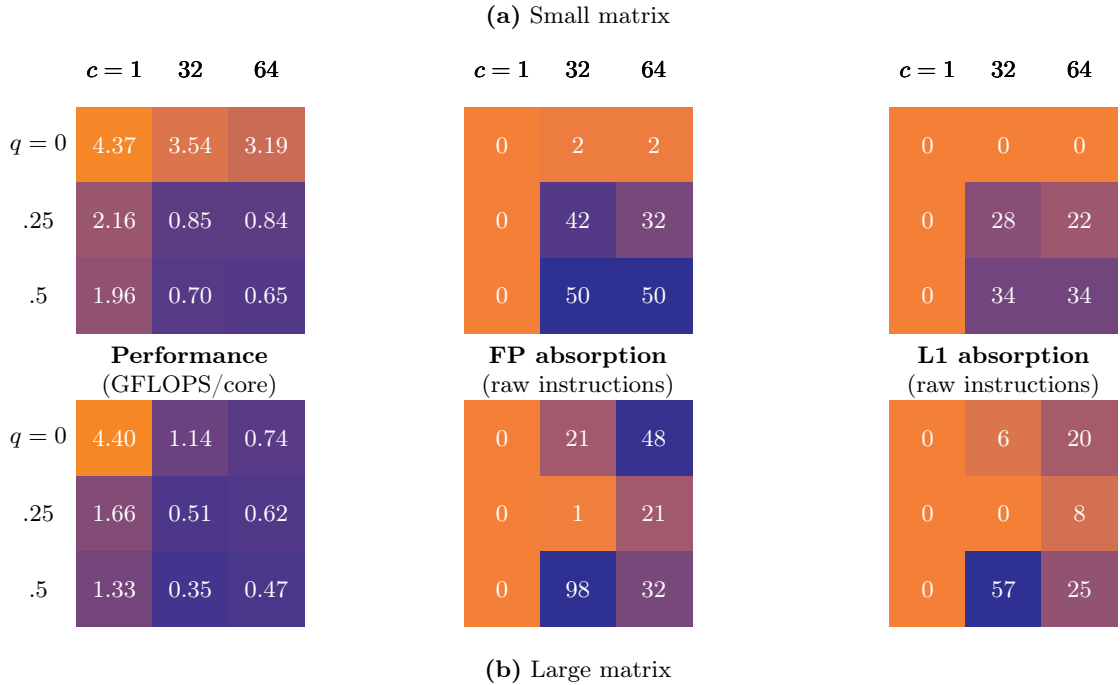
\begin{figure}[tbp]

\centering    \textbf{(a)} Small matrix\\
\begin{minipage}[T]{0.3\textwidth}

\begin{center}

\begin{tikzpicture}[scale=1]
  \foreach \y [count=\n] in {
      {4.37,3.54, 3.19},
      {2.16, 0.85, 0.84},
      {1.96, 0.70, 0.65},
    } {
      \foreach \a [count=\i] in {$c=1$, $32$, $64$} {
        \node[minimum size=10mm] at (\i, 0) {\a};
      }
      \foreach \x [count=\m, evaluate=\x as \gradient using 100-\x*25] in \y {
        \node[fill=Blue!\gradient!Orange, minimum size=10mm, text=white] at (\m,-\n) {\x};
      }
    }
  \foreach \b [count=\j] in {$q=0$, $.25$, $.5$} {
    \node[minimum size=10mm] at (0,-\j) {\b};
  }
\end{tikzpicture}
\\\quad\quad\quad\textbf{Performance}\\\quad\quad\quad(GFLOPS/core)\\
\end{center}
\end{minipage}
\hfill
\begin{minipage}[T]{0.3\textwidth}

\begin{center}
\begin{tikzpicture}[scale=1]
  \foreach \y [count=\n] in {
      {0,2,2},
      {0,42,32},
      {0,50,50},
    } {
      \foreach \a [count=\i] in {$c=1$, $32$, $64$} {
        \node[minimum size=10mm] at (\i, 0) {\a};
      }
      \foreach \x [count=\m, evaluate=\x as \gradient using \x*2] in \y {
        \node[fill=Blue!\gradient!Orange, minimum size=10mm, text=white] at (\m,-\n) {\x};
      }
    }
\end{tikzpicture}
\\\textbf{FP absorption}\\(raw instructions)\\
\end{center}
\end{minipage}
\hfill
\begin{minipage}[T]{0.3\textwidth}

\begin{center}
\begin{tikzpicture}[scale=1]
  \foreach \y [count=\n] in {
      {0,0,0},
      {0,28,22},
      {0,34,34},
    } {
      \foreach \a [count=\i] in {$c=1$, $32$, $64$} {
        \node[minimum size=10mm] at (\i, 0) {\a};
      }
      \foreach \x [count=\m, evaluate=\x as \gradient using \x*2] in \y {
        \node[fill=Blue!\gradient!Orange, minimum size=10mm, text=white] at (\m,-\n) {\x};
      }
    }
\end{tikzpicture}
\\\textbf{L1 absorption}\\(raw instructions)\\
\end{center}
\end{minipage}


\begin{minipage}[t]{0.3\textwidth}

\begin{center}
\begin{tikzpicture}[scale=1]
  \foreach \y [count=\n] in {
      {4.40, 1.14, 0.74},
      {1.66, 0.51, 0.62},
      {1.33, 0.35, 0.47},
    } {
      \foreach \x [count=\m, evaluate=\x as \gradient using 100-\x*25] in \y {
        \node[fill=Blue!\gradient!Orange, minimum size=10mm, text=white] at (\m,-\n) {\x};
      }
    }
  \foreach \b [count=\j] in {$q=0$, $.25$, $.5$} {
    \node[minimum size=10mm] at (0,-\j) {\b};
  }
\end{tikzpicture}
\end{center}
\end{minipage}
\hfill
\begin{minipage}[t]{0.3\textwidth}

\begin{center}
\begin{tikzpicture}[scale=1]
  \foreach \y [count=\n] in {
      {0,21,48},
      {0,1,21},
      {0,98,32},
    } {
      \foreach \x [count=\m, evaluate=\x as \gradient using \x*2] in \y {
        \node[fill=Blue!\gradient!Orange, minimum size=10mm, text=white] at (\m,-\n) {\x};
      }
    }
\end{tikzpicture}
\end{center}
\end{minipage}
\hfill
\begin{minipage}[t]{0.3\textwidth}

\begin{center}
\begin{tikzpicture}[scale=1]
  \foreach \y [count=\n] in {
      {0,6,20},
      {0,0,8},
      {0,57,25},
    } {
      \foreach \x [count=\m, evaluate=\x as \gradient using \x*2] in \y {
        \node[fill=Blue!\gradient!Orange, minimum size=10mm, text=white] at (\m,-\n) {\x};
      }
    }
\end{tikzpicture}
\end{center}
\end{minipage}
\begin{center}
    \textbf{(b)} Large matrix
\end{center}

\caption{Graviton 3 performance (measured in GFLOPS/core) and absorption (FP and L1) of SMPXV on small and large matrices, using varying numbers of cores $c$ and swap probabilities $q$. In matrix (b), it seems counter-intuitive to see both absorptions drop specifically for $q=0.25$. However, this corresponds to the tipping point between the dominance of bandwidth and latency effects.}
\label{figure:spmxvheatmap}
\vspace{-0.5cm}
\end{figure}

SPMXV \footnote{https://git-ce.rwth-aachen.de/hpc-public/epi-spmxv}, a reference benchmark for the European Processor Initiative (EPI), is a sparse matrix-vector multiplication algorithm using Compressed Sparse Row (CSR) storage. This section analyzes the SPMXV kernel on an Amazon Graviton 3 system in light of our methodology. Our goal is twofold:  predict the efficiency of future memory systems (DDR versus HBM) and understand how varying the swapping probability $q$ affects the dominant performance bottlenecks.

The SPMXV kernel accesses matrix elements in a regular, stride-1 fashion, but the vector elements are accessed indirectly based on the matrix’s column indices. By tuning the swapping probability $q$, which randomly swaps non-zero elements within a row, we systematically increase the irregularity of these indirect accesses. This parameter is crucial since it reshapes the memory access pattern at the kernel’s critical multiplication step.
Figure~\ref{figure:spmxvheatmap} shows the performance and absorption levels for varying core numbers $c$ and swap probabilities $q$.
To cover a range of workloads, we consider two different matrices. \textbf{Matrix (a):} A small $134\text{k} \times 134\text{k}$ matrix (44~MB) where, at $q=0$, the entire dataset fits within the L2 and L3 caches. Here, the kernel exhibits excellent scaling with negligible absorption, indicating that performance is primarily limited by core-level effects rather than by memory bandwidth. As $q$ increases, performance degrades and the absorption metric rises, signaling a shift toward latency limitations. \textbf{Matrix (b):} A large $1346\text{k} \times 1346\text{k}$ matrix (480~MB) that, even at $q=0$, shows signs of a memory bandwidth bottleneck with nearly linear access patterns. As $q$ increases, performance declines further; however, the absorption metric first drops and then increases again. This non-monotonic behavior reveals a transition from a bandwidth-bound regime to one dominated by latency effects, a nuance missed by traditional performance measures. Figure~\ref{figure:spmxv_bw2lat} shows how FP absorption captures the two regimes while performance measures do not.

\pgfplotsset{
    scale only axis,
    xmin=0, xmax=0.5,
    scale=0.6
}
\pgfplotscreateplotcyclelist{perfcolors}{Red}
\pgfplotscreateplotcyclelist{abscolors}{Green, Blue}

\pgfplotstableread[col sep=comma]{
q,gflops
0.0,0.76
0.1,0.69
0.2,0.64
0.25,0.62
0.3,0.59
0.4,0.52
0.5,0.47
}\perfdata

\pgfplotstableread[col sep=comma]{
q,abs
0.0,48
0.1,4
0.2,13
0.25,25
0.3,31
0.4,32
0.5,32
}\fpabsdata

\begin{figure}[tbp]
\begin{minipage}[t]{0.55\textwidth}
{
\begin{tikzpicture}

    \begin{axis}[
    axis y line*=left,
    ymin=0, ymax=0.8,
    xlabel={Swapping probability $q$},
    scaled ticks=false,
    ylabel={Performance [GFLOPS/core] \ref{pgfplots:gflops}},
    cycle list name=perfcolors
    ]
    \addplot[color=Red, mark=none, dashed, line width=0.40mm] table[x=q,y=gflops] {\perfdata};
    \label{pgfplots:gflops}
    \end{axis}

    \begin{axis}[
        axis y line*=right,
        axis x line=none,
        ylabel style={align=center},
        ylabel={Raw fp\_add64 absorption \ref{pgfplots:fpabs}},
        cycle list name=abscolors
    ]
    \addplot[color=Green, mark=none] table[x=q,y=abs] {\fpabsdata};
    \label{pgfplots:fpabs}
    \end{axis}

\end{tikzpicture}
}
\centering
\caption{Evolution of performance and absorption on large matrix (b) in function of swapping probability $q$ (G3, 64 cores). While performance only decreases, absorption drops and increases again, indicating a transition between two different regimes (presumably bandwidth, then latency boundness).}
\label{figure:spmxv_bw2lat}
\vspace{-1cm}
\end{minipage}
\hfill
\begin{minipage}[T]{0.35\textwidth}

\begin{center}
\captionof{table}{Performances of SPMXV on large matrix (b) running on Sapphire Rapids. HBM performance collapse when $q$ increases, confirming that with higher values of $q$ the benchmark becomes latency-bound.}
\begin{tabular}[t]{l p{1.3cm} p{1.3cm}}
\toprule

\textbf{Performances} & DDR & HBM  \\
(GFLOPS/core) && \\
\midrule
\midrule

$q=0$ & 0.239 & 0.238 \\
$q=0.25$ & 0.233 & 0.066\\
$q=0.5$ & 0.201 & 0.058\\

 \bottomrule  
\end{tabular}

\label{table:spmxv_ddrhbm}

\end{center}

\end{minipage}

\end{figure}

To further confirm our regime-transition hypothesis, we ran the large matrix on a Sapphire Rapids Intel Xeon Max~9460 system with both DDR and HBM memory. While performance at $q=0$ was similar for both memory types, higher $q$ values resulted in a dramatic collapse in performance on HBM. This supports our analysis: instead of fetching small amounts of data frequently, HBM retrieves large bursts of data at once, which has a severe penalty when random memory accesses become prevalent. Notably, conventional tools fall short in this analysis. DECAN, for example, struggles because it duplicates the innermost loop (which typically iterates only a few times) leading to unstable timing measurements. Static models (e.g. MAQAO CQA/UFS) do not apply here, as the kernel is dominated by cache and memory effects. Hardware counters yield multiple, often confusing metrics that obscure the underlying bottleneck. In contrast, our noise injection approach provides a single, intuitive absorption metric that was able to pinpoint the transition between bottleneck regimes. Noise injection captures the subtle interplay between bandwidth and latency limitations in the SPMXV kernel. This capability to detect hidden performance regimes underscores the value of our tool in scenarios where traditional methods are either inapplicable or overly complex to interpret and provide actionable insights to select between different memory families.

\section{Conclusion and Future Work} \label{section:conclusion}

\noindent\textbf{Contributions.}
In this paper, we introduce a model-agnostic, instruction-accurate noise injection framework for bottleneck analysis. Our approach inserts carefully selected assembly instructions into performance-critical loops to probe specific hardware resources, namely the floating-point unit, L1 cache, and memory, to quantify the available noise through an absorption metric. By controlling the injected noise, we determine the precise point at which performance begins to degrade, allowing us to classify regions as compute-, bandwidth-, or latency-bound. We validate our framework on a suite of hardware benchmarks, demonstrating its effectiveness for comparing different systems and its advantages over methods that rely on oversimplified assumptions. By doing so, we notably show that fine-grained unicore optimizations are often overshadowed by the bandwidth or latency of memory accesses that appear in more representative multicore scenarios. Moreover, our application to the SPMXV benchmark shows that the absorption metric reliably captures transitions between bandwidth and latency limitations and offers practical insights for selecting between HBM or DDR memory.\\

\noindent\textbf{Future Work.}
Future efforts will focus on the continuation of the presented experiments, as well as methodological refinements and tool enhancements. Notably, experiments on lat mem rd and SPMXV suggest that it is not easy to characterize how latency stalls can be exploited, and different access patterns seem to create complex behaviors that should be further explored. On the implementation side, we plan to streamline the user workflow by automating the identification of hot loops. This can be done by instrumenting all loop nests at the IR level so that users would no longer need to separately profile the target code. Methodologically, our framework permits the definition of additional noise modes; future studies will explore more complex and combined patterns and extend noise injection to target other resources, such as intermediate cache levels or I/O subsystems. Finally, selectively injecting noise into specific threads or processes during parallel communications may provide deeper insights into applications' resilience to desynchronization at both the thread and node levels.

\medskip\noindent Source code of the LLVM injection plugin and information to reproduce this paper's experiments can be found on the following GitHub repository: \url{https://www.github.com/sipearl/eris}

\medskip\noindent\textbf{Disclosure of Interests.}
The authors have no competing interests to declare that are relevant to the content of this article.


\bibliographystyle{splncs04}
\bibliography{bibliography}

\end{document}